\documentclass[sigconf]{acmart}

\AtBeginDocument{%
  \providecommand\BibTeX{{%
    \normalfont B\kern-0.5em{\scshape i\kern-0.25em b}\kern-0.8em\TeX}}}

\setcopyright{acmcopyright}
\copyrightyear{2023}
\acmYear{2022}
\acmDOI{10.1145/xxxxxxx.xxxxxxx}

\acmConference[ACSAC '23]{ACSAC '23: Annual Computer Security Applications Conference}{December 4–8, 2023}{Austin, TX, USA}
\acmBooktitle{ACSAC '23: Annual Computer Security Applications Conference, December 4–8, 2023, Austin, TX, USA}
\acmPrice{15.00}
\acmISBN{978-1-4503-XXXX-X/21/06}

\newcommand{\TagModelOne}{BLE-AC}
\newcommand{\TagModelTwo}{UWB-RAW}
\newcommand{\NumberOfParticipants}{23}
\newcommand{\NumberOfParticipantsUndergraduates}{16}
\newcommand{\NumberOfParticipantsGraduates}{7}
\newcommand{\NumberOfParticipantsUndergraduatesSTEM}{six}
\newcommand{\NumberOfParticipantsGraduatesSTEM}{four}
\newcommand{\ApplicationName}{DIAL}
\usepackage{subcaption}
\begin{document}

\title{A Tagging Solution to Discover IoT Devices in Apartments}

\author{Berkay Kaplan}
\authornote{Both authors contributed equally to this research.}
\email{berkayk2@illinois.edu}
\orcid{0000-0002-4365-7606}
\affiliation{%
  \institution{University of Illinois Urbana-Champaign}
  \streetaddress{P.O. Box 1212}
  \city{Champaign}
  \state{Illinois}
  \country{USA}
  \postcode{61820}
}

\author{Jingyu Qian}
\authornotemark[1]
\email{jingyuq2@illinois.edu}
\orcid{0000-0002-3953-5382}
\affiliation{%
  \institution{University of Illinois Urbana-Champaign}
  \streetaddress{P.O. Box 1212}
  \city{Champaign}
  \state{Illinois}
  \country{USA}
  \postcode{61820}
}

\author{Israel J Lopez-Toledo}
\email{israell2@illinois.edu}
\affiliation{
  \institution{University of Illinois Urbana-Champaign}
  \streetaddress{P.O. Box 1212}
  \city{Champaign}
  \state{Illinois}
  \country{USA}
  \postcode{61820}
}

\author{Carl Gunter}
\email{cgunter@illinois.edu}
\affiliation{
  \institution{University of Illinois Urbana-Champaign}
  \streetaddress{P.O. Box 1212}
  \city{Champaign}
  \state{Illinois}
  \country{USA}
  \postcode{61820}
}

\renewcommand{\shortauthors}{Kaplan et al.}

\begin{abstract}
The number of Internet of Things (IoT) devices in smart homes is increasing. This broad adoption facilitates users' lives, but it also brings problems. One such issue is that some IoT devices may invade users' privacy through obscure data collection practices or hidden devices. Specific IoT devices can exist out of sight and still collect user data to send to third parties via the Internet. Owners can easily forget the location or even the existence of these devices, especially if the owner is a landlord managing several properties. The landlord-owner scenario creates multi-user problems as designers typically build IoT devices for single users. We developed tag models that use wireless protocols, buzzers, and LED lighting to guide users toward the hidden device in shared spaces and accommodate multi-user scenarios. They are attached to IoT devices inside a residential unit during their installation to be later discovered by a tenant. These tags are similar to Tile models or Airtag but have different features based on our privacy use case. For instance, our tags do not require pairing; multiple users can interact with them through our Android application. Our tags can also embed the IoT device's information while protecting against unwanted access to that information through a proximity requirement. Researchers have developed several other tools, such as thermal cameras or virtual reality (VR), for discovering devices, but we focused on wireless technologies. We measured specific performance metrics of our tags to analyze their feasibility for this problem. We also conducted a user study to measure the participants' comfort levels while finding objects with our tags attached. Our results indicate that wireless tags can be viable for device tracking in residential properties.
\end{abstract}



\maketitle

\section{Introduction}
Smart homes use IoT devices to improve the occupants ' lives. As of 2021, approximately 43\% of households in the U.S. own a smart device, increasing from 33\% in 2019 \cite{vailshery_2021}. Commonly available IoT devices, such as learning thermostats, video doorbells, smart baby monitors, and voice-controlled devices, are also relatively affordable for any consumer \cite{zheng2018user}. As IoT devices' popularity increases and they upload more data to the cloud, privacy questions, such as data collection practices, arise \cite{zheng2018user, castelli2017happened}. For instance, although voice assistants only activate when they hear specific keywords, their implementation requires them to listen to their environments constantly \cite{chen2020wearable}. They can start recording conversations maliciously or by misconfiguration \cite{chen2020wearable}. In addition, IoT devices contain several design flaws and vulnerabilities that may have devastating consequences on users' privacy \cite{shen2019iot,colnago2020informing}. These design flaws may cause sensitive information, such as conversation recordings, to leak onto the Internet. 

These privacy questions created a new branch that researchers attempt to understand: intelligent home IoT devices in multi-user scenarios \cite{geeng2019s,marky2020don}. IoT devices in shared environments become shared devices that affect multiple people \cite{geeng2019s, zeng2019understanding}. Nevertheless, it is crucial to make home data more accountable in shared settings \cite{castelli2017happened}. Furthermore, some popular IoT platforms might not comprehensively address multi-user scenarios \cite{zeng2019understanding}. For instance, a question thread created in the SmartThings community forum in 2017 indicated that end-users could create multiple accounts. However, they cannot give the accounts different access levels to information \cite{smartthingscommunity_2017}. The lack of shared-space settings in IoT devices can further exacerbate privacy concerns.

Such environments affected by this inadequacy are rental apartments, such as Airbnb, and hotels, which all concerned researchers \cite{dey2020exploring}. Especially, hidden devices in these properties can make users uncomfortable \cite{emami2019exploring}. For instance, a computer science professor in an Airbnb rental found a camera that views a field close to the bathroom, according to the Washington Post \cite{song2020m,shaban_2019}. Tenants unknowingly living with hidden IoT devices may be victims of privacy violations \cite{yao2019defending}. The landlord also may not remember the location and information of each installed device as they may have several apartments. Tenants need an automatic solution to alert them of each IoT device's existence. 

We offer to tag each device inside the apartment. The primary purpose of these tags is to allow tenants to discover, locate, and inventory each device inside a room via wireless protocols, buzzers, and LED lights. The tags will alert each user nearby that a device exists. Thus, users will learn the location and the device information using wireless capabilities. With this transparent mechanism, user privacy is respected, giving users a choice to leave the apartment or shared space. They can also opt to contact the owners for the device's removal.

Our solution contains two tag models that consist of small circuit boards that utilize various wireless tools to interact with an Android application named \ApplicationName. It is an interface for users to discover every tag nearby and identify and inventory the device. This identifying information can be a web link pointing to the device details, such as an Amazon sales page. The user can Google further information regarding its data collection practices. Another alternative is to point to the product's manufacturer privacy page to prevent users from spending effort. We gave the freedom for tag administrators to decide what to store in them.

We used relatively cheap and publicly available circuit boards from vendors, including Qorvo and Adafruit. These boards use Bluetooth Low Energy (BLE), Ultra-wideband (UWB), and LED lights. We also used coin-sized tags for Near-field communication (NFC) as the medium to transmit device information and a buzzer to enable tags to make noise to reveal their location. The LED lights present in both models also help users locate the device if the tag is visible. We built different tag models to provide users with the cheapest and longest-ranged tags. Our tag models do not require initial pairing and can attach to devices with all communication protocols. To our knowledge, the wireless device discovery and identification solution has not been proposed.

Besides discussing our model implementation, we also conducted a user study to understand potential users' perceptions of our tag models. We conducted trials where we asked participants to find our tags attached to things using \ApplicationName{} in an apartment. Afterward, the participants took the System Usability Scale (SUS) survey, a quick method to evaluate the usability of human-machine systems \cite{peres2013validation}. Finally, we discussed the feasibility of our solution concerning user perception and technical facts. Our primary and novel contributions to the literature are:
\begin{itemize}
  \item a tagging implementation for discovering and identifying hidden IoT devices,
  \item a user study on the tags that focuses on participants' comfort levels and price acceptability,
  \item an analysis of our wireless tagging solution's feasibility.
\end{itemize}

We organize our paper as follows. In the background section, we will first explain some of the wireless protocols we are using to give readers an understanding of the implementation of the tags. The model overview section will overview the total model and the solution's workflow. The implementation section will describe the electronic components and algorithms we use in our system. In the evaluation section, we will present our user study and evaluate the tags' performance. The discussion section will contain the tagging feasibility analysis and possible future tracks to extend this project. Then, we will present related work from the literature focusing on device discovery. Finally, we summarize the key points of this paper and describe future predictions using wireless for device discovery.

\section{Background}
We used several wireless protocols, some of which may not be familiar to readers. Thus, it would be helpful to give a background on these protocols.

We used NFC to identify the device our tag is attached to, as it will hold a link that points to a web page. NFC is a wireless protocol that allows users to transfer information between a tag to a reader using NFC Data Exchange Format (NDEF) \cite{want2011near}. It operates at 13.56 MHz and has a data rate of 424 kbit per second \cite{want2011near}. The range is only 10 centimeters and supports data transfer between two readers. NFC is a great candidate for our project to limit physical access to sensitive information.

We also needed a protocol with a range suitable for a residential unit's room while preserving power. One such tool we found is the BLE, which consumed less than one milliampere during our trials. BLE is a complementary technology to Bluetooth Classic and borrows several techniques from its parent tool while having completely different goals and market segments. BLE optimized its power consumption for ultra-low power rather than focusing on increasing its data rate \cite{heydon2012bluetooth}. Although designers intended BLE to work with coin-cell batteries, we want our BLE tag to have a battery life of more than a year \cite{heydon2012bluetooth}. Due to its ultra-low power consumption and configurable ranges through transmitter power, it is a good option for our design. A buzzer would allow the user to find its location in out-of-sight cases with a sound.

A more advanced method than a buzzer making noise for location tracking is UWB. It aims to provide a low-complexity, low-cost, low-power consumption, and high data-rate wireless alternative in personal ranges \cite{aiello2003ultra}. The Federal Communications Commission (FCC) allocated 7,500 MHz of spectrum for the unlicensed use of UWB in the 3.1 to 10.6 GHz frequency band. UWB's range of 12 feet was suitable for residential environments, although the power consumption of our boards was high for our batteries \cite{aiello2003ultra}. Nevertheless, we still decided on this tool as a discovery and location tracking method. Some residential properties may have noise in certain rooms. Offering another solution for a noisy environment would enhance the tags' use cases.

\section{Model Overview}
\begin{figure}
    \centering
    \includegraphics[scale=.3]{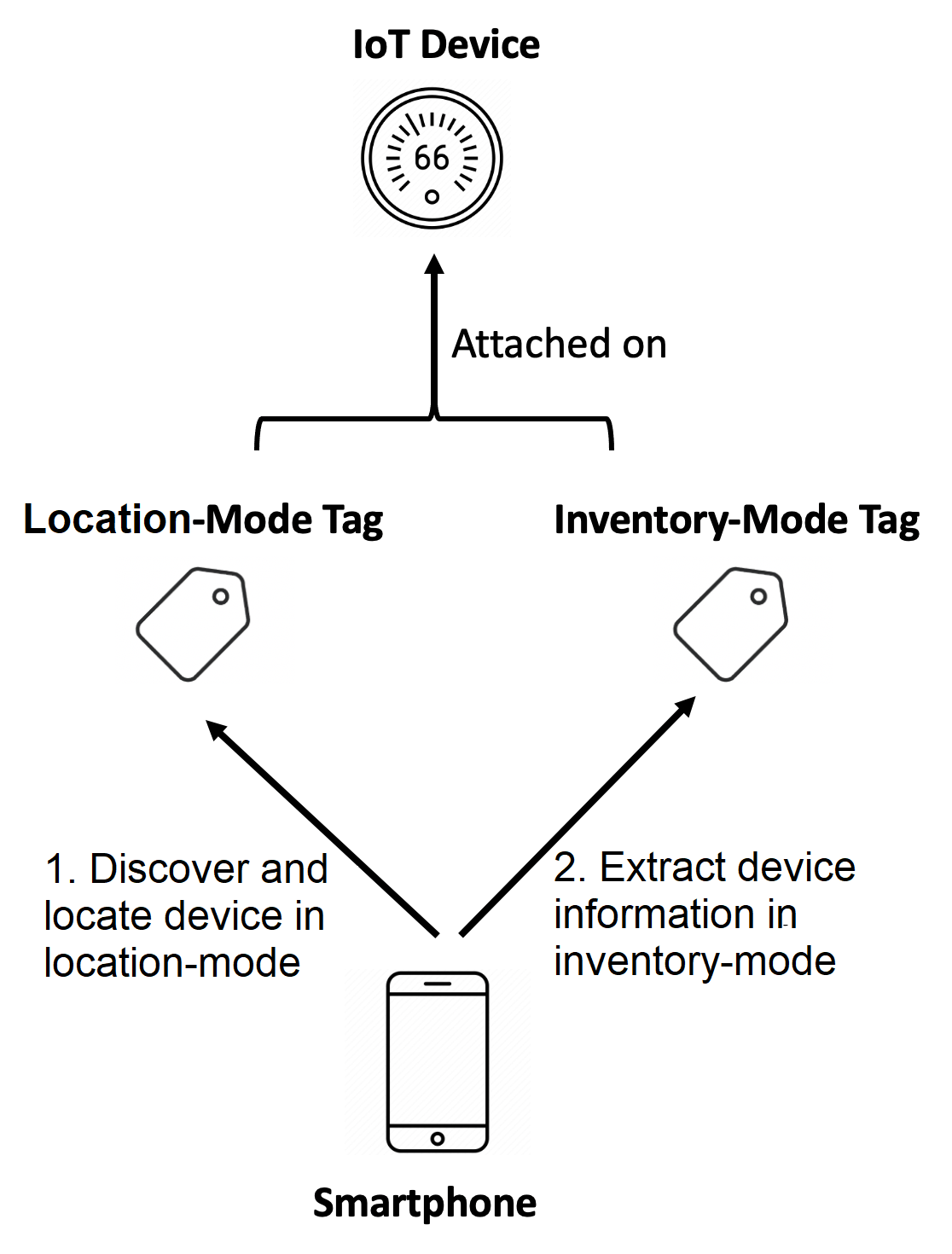}
    \caption{Two-tag model for location and inventory modes.}
    \label{fig:model}
\end{figure}

This section discusses our model designs to discover and locate hidden IoT devices and extract their information (Fig. \ref{fig:model}). We divide the two steps into location mode and inventory mode because of the different purposes of each step and privacy requirements. In the location mode, the user tries to discover and locate the hidden device. The information the user gains in location mode will only indicate the existence and location of an IoT device, as it will not contain valuable data. In contrast, the user intends to learn detailed information about the discovered device in the inventory mode, such as its vendor and software versions. The smartphone app works as the user interface to guide the user to the IoT device via sounds or UWB. Once the user is within reach of the tag, he can scan the NFC tag to display detailed information. Finally, he can inventory all the IoT devices he found in the apartment.

\subsection{Threat Model}

Our design assumes that the landlord or the device manufacturer is collaborative. Collaborative landlords are willing to attach the tags to the IoT devices. This assumption is reasonable because tags are cheap, and landlords want to keep tenants comfortable to avoid bad reviews and potential privacy infringements. Collaborative device manufacturers may also want to improve public relations by implementing privacy-friendly product solutions. 


The tags do not require configuration except for the NFC coin holding information about the attached thing. We assume the landlord or manufacturer loads this URL into the NFC coin. With this setup, the renter can identify the things with \ApplicationName. 

If the renter is tech-savvy, he can still configure the NFC coin with desired extra information through his mobile phone using a third-party NFC reader-writer application. However, we do not consider malicious renters. In everyday scenarios, renters can plant pinhole cameras, change the information inside the NFC coin, or perform malicious operations. For the scope of our design, we assume renters are not evil. 

We assume that our tag models are only intended for those smart home devices inside the apartment. Our targeted users are the residents of the apartment. We assume they are more concerned about what those devices are and how they collect personal data in their environment. Temporary guests interacting with the tags is the tenant's responsibility and is an out-of-scope scenario for our project.

\subsection{Device Discovery and Locating}

Tags are attached to IoT devices and broadcast signals to the smartphone via BLE or UWB. Then, \ApplicationName \hspace{0pt} guides the user to the IoT device. When \ApplicationName \hspace{0pt} is in the location mode screen, it will display all the tags in approximately 15 - 20 meters. The user can buzz each tag individually or find the distance to the tag via UWB. These BLE beacons and UWB signals contain no information regarding the attached thing to prevent privacy leakage. Thus, an attacker can only gain the number of tags in an environment if he has \ApplicationName{} or knows our beacon formatting. In addition, since the tags do not contain an authentication method for activating the buzzer and UWB, the attacker can also buzz his neighbor's tags or find the tag's location via UWB. Future work can include developing a more advanced formatting methodology to avoid this issue. Another theoretical solution would be that the NFC tag can hold a password that activates the buzzer. Since the NFC has a proximity requirement, only the resident can access this password. Without the password, the buzzer feature would be dormant. 

\subsection{Device Information Extraction}

Once the user finds the hidden IoT device, he needs to learn about the device to be more familiar with the smart home environment. He must also educate himself regarding the device vendor or any third parties using his private data. Due to the privacy level of this information, the user should be the sole person to extract this information. Therefore, we rely on NFC to transmit device information. We integrated an NFC tag into our tag models so the smartphone equipped with NFC capability can easily extract device information. This design also guarantees the proximity requirement if anyone wants to read from the tag. 

Users can place several types of data in the NFC tag. Basic IoT device information, such as the device name, functionalities, and vendors, should be put in the NFC tag first. Moreover, data collection activities are critical to helping users make better security and privacy decisions. Finally, some extra information, such as firmware versions and software vulnerability histories, can inform users whether updates are required to guard against the latest attacks. All such critical device information can exist on a single NFC tag.

\subsection{Tag Models}

Listening to and broadcasting BLE beacons is an inexpensive protocol in terms of battery life. Our BLE boards range approximately 15 meters, which we found optimal in an apartment setting. Therefore, we decided to use it to detect the device's existence and a buzzer to locate it. On the other hand, our UWB boards provide more accurate device locations and have a range of 5 meters. Since UWB and BLE have similar functionalities in the location mode, we did not combine them in one tag model. We decided to offer them separately to give users a choice, as UWB can provide exact locations even in noisy environments. However, BLE also has the advantage of low power consumption and a higher range. We wanted to give our users a choice based on their needs.

We design two tag models for users. Our first model uses BLE with a buzzer attached. The model uses BLE to achieve a coarse tag location through received signal strength indication (RSSI), while the buzzer can provide a more accurate location service. We named this model \TagModelOne. Our second model uses UWB. This model allows us to achieve a precise device location within the apartment due to UWB's strong indoor positioning capability. We named this model \TagModelTwo. We use a coin-size NFC tag for the inventory mode in both tag models to ensure that only the nearby reader can extract the device information. Both tag models have LEDs that the user can activate to locate the tags once they become visible to the user.

\subsection{Tag Reader}

\begin{table*}[ht]
\centering
\caption{Comparison of our tags to other popular device discovery and inventory solutions.}
\label{tab:comp}
\begin{tabular}{ccccc}
  \hline
  {\bfseries Technology} & {\bfseries No pairing} & {\bfseries Network protocol unrestricted} & {\bfseries Support locating device} & {\bfseries Inventory device information}\\
  \hline
  Privacy Labels & \checkmark & \checkmark & & \checkmark\\
  Airtags and Tiles & & \checkmark & \checkmark & \checkmark \\
  SmartThings & & & & \checkmark\\
  Lumos & \checkmark & & \checkmark & \\
  Our Tags & \checkmark & \checkmark & \checkmark & \checkmark\\
  \hline
\end{tabular}
\end{table*}

One of the primary considerations when selecting a tag model is its reader's price. Our current design utilizes BLE, NFC, and UWB tags. For BLE and NFC, we can rely on smartphones as tag readers. Almost all smartphones today support Bluetooth and BLE. In addition, 73\% of smartphones in 2018 support NFC~\cite{UWB_support}. On the other hand, UWB used to be an expensive technology, but now it is cheaper. Few smartphones now have UWB, including iPhones after the iPhone 11 and Samsung Galaxy S21 series. We have an adapter with UWB capabilities for older smartphones that can connect to the phone through the USB serial port, working as the UWB tag reader. This UWB reader can communicate with the tag. We then built \ApplicationName \hspace{0pt} to read the distance from the tag reader through the Android USB serial port, which provides the distance in meters to guide the user to the IoT device. Although our current design requires an external adapter for older smartphones, we expect smartphone companies will incorporate UWB in their following models as manufacturers phase out older phones.

\subsection{Comparison of Our Tags to Existing Solutions}

In this section, we show the novelty of our tag models by comparing them with existing solutions for IoT device discovery and inventory, including privacy labels, commercial tags (e.g., Airtag~\cite{apple} and Tiles~\cite{tile}), the smart home manager (e.g., SmartThings~\cite{smartthings}), and research work (e.g., Lumos~\cite{sharma2022lumos}). Our comparison is from two angles: utility and privacy. 

\subsubsection{Utility Analysis}
We compare our tag models with the techniques mentioned above based on the following features (Table \ref{tab:comp}):
\begin{itemize}
    \item Does the technology require initial pairing?
    \item Is the technology designed for devices with some specific network protocol (e.g., WiFi devices)?
    \item Can the technology locate the device?
    \item Can the technology reveal device information to the user? 
\end{itemize}

The privacy labeling scheme conveys high-level security and privacy facts regarding the IoT device to users to raise their privacy awareness \cite{shen2019iot}. Our tags have similar purposes of privacy labels, but Shen et al. \cite{shen2019iot} did not provide a medium to transmit this label to the users. However, we can integrate the contents of the privacy labels from this work with our tags. Our tags can hold a certain amount of information, such as web links, leading to a privacy label page. Privacy labels do not require initial pairing and are not restricted by the device's communication protocol. It can be in paperback or transmitted via any wireless tool. Although they do not support locating the device, they still enable the user to learn device information.

Airtag from Apple and Tile models are commercial solutions for problems including lost item tracking, but they can still be attached to home IoT devices after being paired with the tenant's phone. We design our tags to fit multi-user scenarios more and provide a solution for device discovery in a cooperative environment where the manufacturer, vendor, or landlord helps implement our tags. Our tag models cover the multi-user scenario by removing pairing and allowing any tenant to download \ApplicationName \hspace{0pt} to discover nearby devices. Nevertheless, Airtags and Tiles still provide accurate locating and can hold information regarding the device it is attached to. It is also independent of the attached device's communication protocols, as each tag can be attached to any device, even if it does not connect to the Internet.

SmartThings is a home automation platform developed by Samsung for device discovery that would enable users to control nearby IoT devices. Although it uses a broad range of wireless protocols, including ZigBee and Z-Wave, it does not cover a comprehensive set of communication protocols. Users must add compatible devices to SmartThings or pair them with the hub. Although some SmartThings compatible devices have location-tracking features, SmartThings, by default, does not help locate the device but can inventory and manage device information. However, not all devices are compatible with SmartThings. Our tags can be attached to any device, even if they do not connect to the Internet, are in sleep mode, or have not been turned on. 

Lumos~\cite{sharma2022lumos} utilizes wireless traffic monitoring to identify and locate hidden WiFi-connected IoT devices in public locations, such as hotels and Airbnb. It is designed only for WiFi-connected devices and requires them to be on so that it can sniff the ongoing traffic. Therefore, Lumos~\cite{sharma2022lumos} does not need initial pairing to locate the devices. However, Lumos~\cite{sharma2022lumos} does not transmit or inventory device information. 

Compared to existing solutions, our tag models fulfill all the features in Table \ref{tab:comp}. Our tag models resemble popular tracking tags, such as Airtags and Tile. However, these solutions mainly track lost items and do not provide enough information about the attached object, except for metadata, such as a name or picture. Therefore, they could only partially solve the device discovery problem. In addition, users must pair the tracking tag and their mobile phones each time. The requirement of initial configuration for permission to interact with the tag is not well-suited for multi-user scenarios. Every shared space or apartment will have temporary tenants, and it would be inconvenient for each tenant to find every hidden device, possibly hidden inside the walls, and pair it with his phone for his stay or lease.

\subsubsection{Privacy Analysis}
Here, we illustrate the advantage of our tag models in terms of privacy compared to Airtag and Tiles.

Airtag and Tiles can be exploited for malicious purposes, such as spying on someone. Although they provide a certain level of protection, a journalist's experiment showed that the victim could not find the Airtag, although he received a notification of its presence \cite{hill_heisler_2022}. Nevertheless, Airtags can still play sounds if a tracking victim cannot find it \cite{apple}.

Since Airtags communicate with billions of phones worldwide to track their locations, they are much better informed than our tags or the Tiles. However, if an iPhone detects an Airtag continuously moving with it, the iPhone alerts its owner with a notification \cite{hill_heisler_2022}. The notification includes the entry point where the tracking started. Tiles do not have this feature, while Airtags are too well-informed. The intensive iPhone network for location tracking from Airtag and the lack of spying protection from Tiles do not make them suitable for privacy-sensitive use cases.

Our tag models are designed to protect IoT user privacy in a multi-user setting. We identify different utility and privacy requirements in various stages of IoT device discovery. We need a wireless protocol with a reasonable range covering the typical apartment space to locate the device. Therefore, we attach the BLE or UWB tag to the device. Unlike Airtag, we do not use crowdsourcing techniques to help find the device, which we believe to be overkill for a multi-user apartment setting and can open more attack vectors. To prevent privacy leakage due to BLE or UWB's range, we carefully keep only the required information in the beacon to guide the user to the device. Any detailed information related to the device is never sent out. On the other hand, we assume that proximity can be used as an authorization method to access the IoT device. Therefore, we use an NFC tag with a minimal range to transmit detailed device information to the user. This design allows only the user near the device to gather such private information.

\section{Implementation}
This section will discuss the specific boards and techniques used for the tag models. We will also present an overview of \ApplicationName \hspace{0pt} to interact with the tags.

\subsection{Tag Implementations}

Our two tag models include electronic boards, a buzzer, and a coin-sized NFC tag. The first tag model, \TagModelOne, is built with an Adafruit ItsyBitsy nRF52840 Express \cite{industries}. This board also serves as the microcontroller to broadcast BLE beacons periodically and activate the buzzer when the user signals. The LED light will also be active while the buzzer makes a noise. The board will broadcast a beacon A with a specific formatting every half a second and simultaneously listen to beacons in the area for the buzzer and LED activation. The beacon the board accepts will be the same as beacon A, so only one tag's buzzer will be activated.

Our Android application will first receive the board's beacon. It will recognize this beacon as its data has a specific formatting type. Afterward, the app will broadcast it back to activate the board's buzzer when the user prompts \ApplicationName. The particular formatting is a simple method, as the sum of every hex value in the beacon will yield a specific value. A more advanced way can avoid security issues, such as buzzing strangers' tags. The price of this board is only \$19.95 \cite{industries}.

We found the Adafruit ItsyBitsy nRF52840 Express board easier to program. The board uses Arduino IDE to upload code, and we found rich documentation online. We also attached the buzzer with two pins to the Adafruit board. The first pin is connected to one analog output pin, and the second to the ground pin. This analog pin outputs pulse-width modulation signals with variable frequency. This frequency-varied signal allows the buzzer to make sounds at different frequencies. The bottom of the board houses the NFC tag via an attachment tool, such as duct tape.

The second tag model, \TagModelTwo, is built with the \$19.50 DWM1001-DEV board from Qorvo \cite{symmetry}. This board has the UWB feature and BLE for configuration purposes \cite{mouser_2020}. We found this board to be more challenging to program with SEGGER. Nevertheless, we completed our action items for this board as we found that Decawave offered a handful of example programs on GitHub \cite{decawave}. The only issue is that the DWM1001 design is an anchor point where designers assume it can access unlimited power \cite{nudel_2019}. Thus, its developers did not focus on power consumption optimization \cite{nudel_2019}. Therefore, the battery life will be only a few days, even with three AA batteries.

\begin{figure}[h]
    \centering
    \includegraphics[width=0.6\columnwidth]{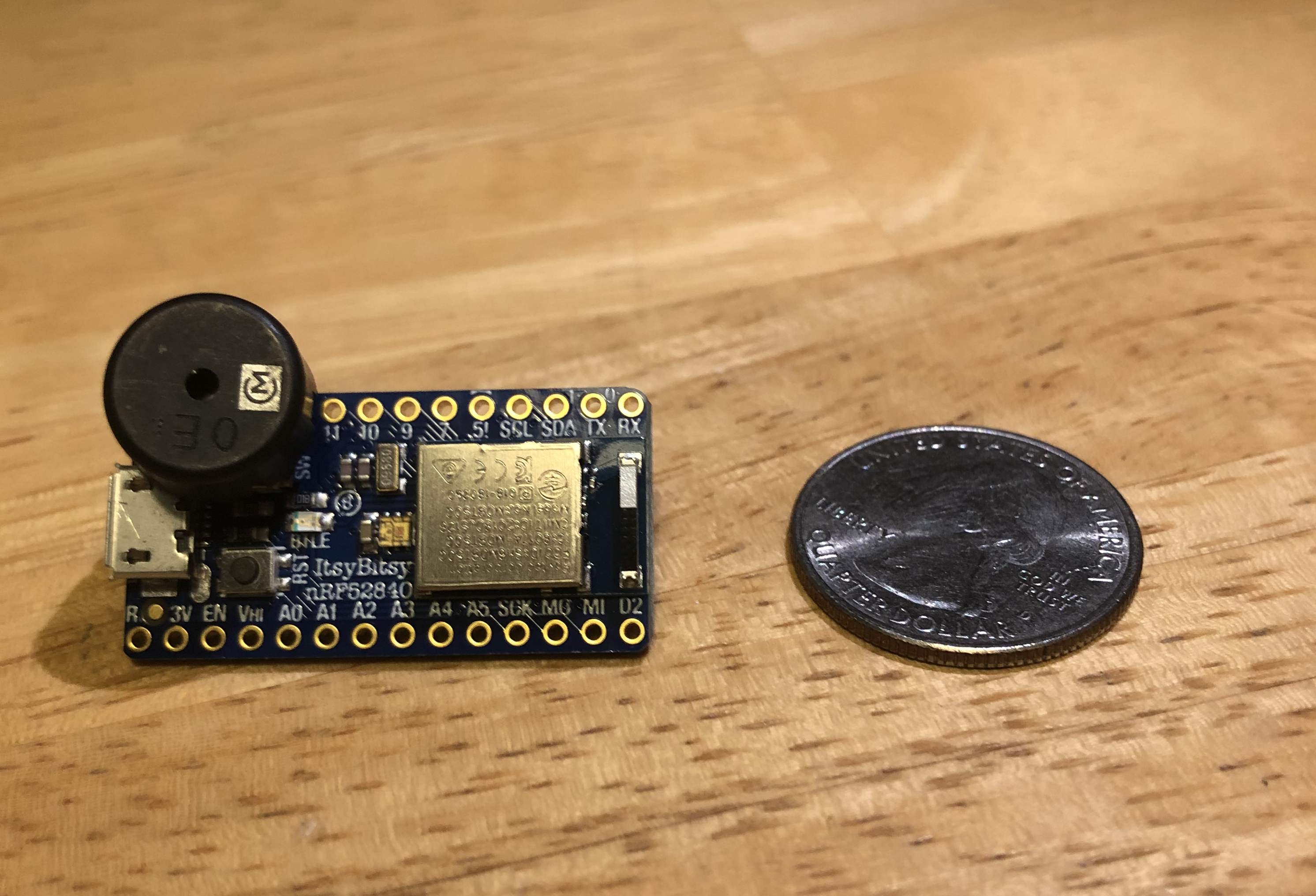}
    \caption{The size comparison of an Adafruit ItsyBitsy nRF52840 Express and the buzzer to an American quarter.}
    \label{figure:ItsyBitsyImage}
\end{figure}

\begin{figure}[h]
    \centering
    \includegraphics[width=0.6\columnwidth]{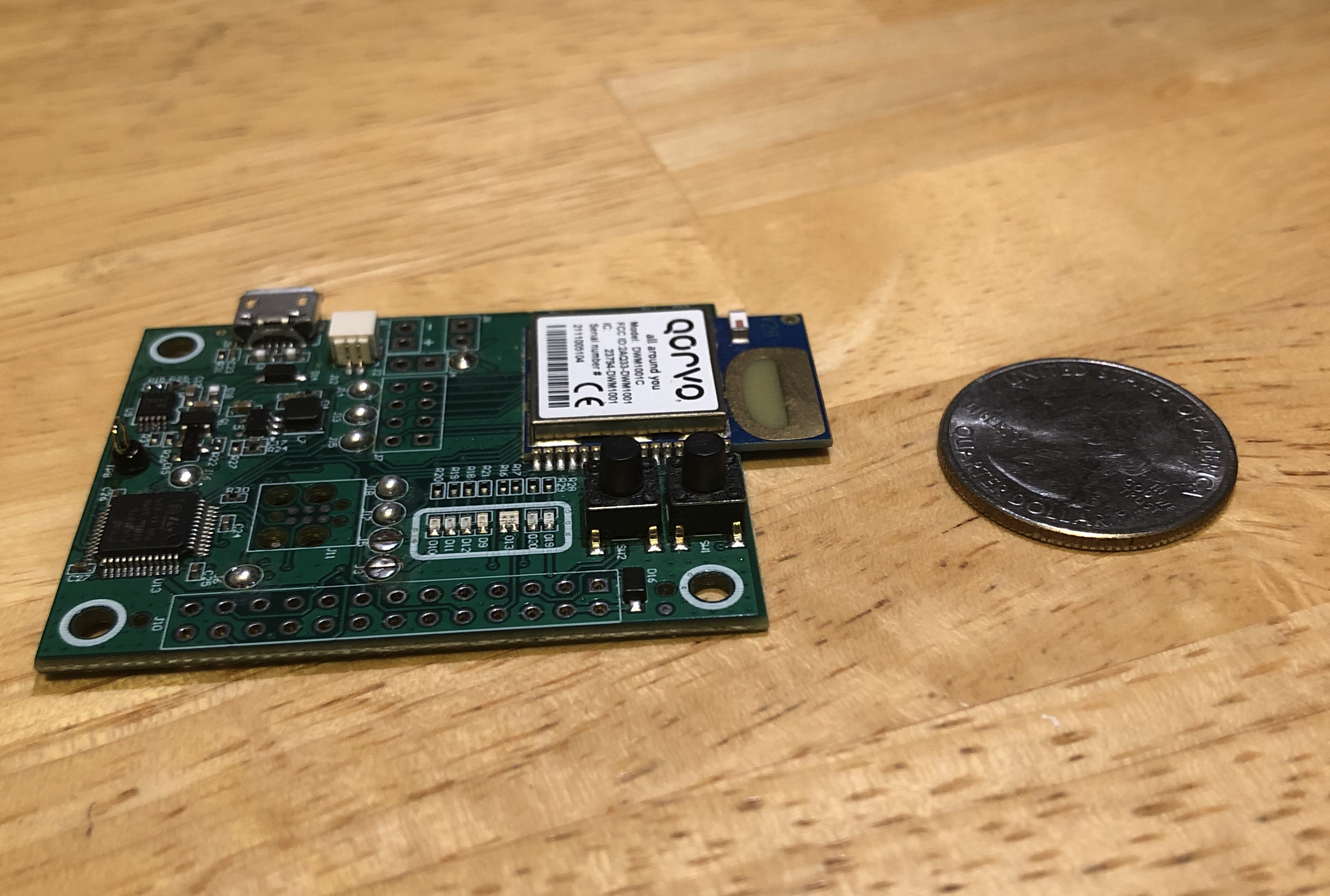}
    \caption{The size comparison of a DWM1001-DEV to an American quarter.}
    \label{figure:DWM1001Image}
\end{figure}

Once a phone with UWB enabled comes closer, it will start receiving the distance in meters through \ApplicationName. This distance is between the tag and the phone with our app. The user can determine a direction to move towards that continuously decreases the length. The LED light will always be on, helping the user locate the tag once he gains visual contact. The DWM1001 development board also has the NFC tag attached at the bottom, similar to the first tag model. Both these tag models are in Fig. \ref{figure:ItsyBitsyImage} and \ref{figure:DWM1001Image}.

\begin{figure*}[h]
\centering
\begin{subfigure}[b]{0.26\textwidth}
\includegraphics[width=\textwidth]{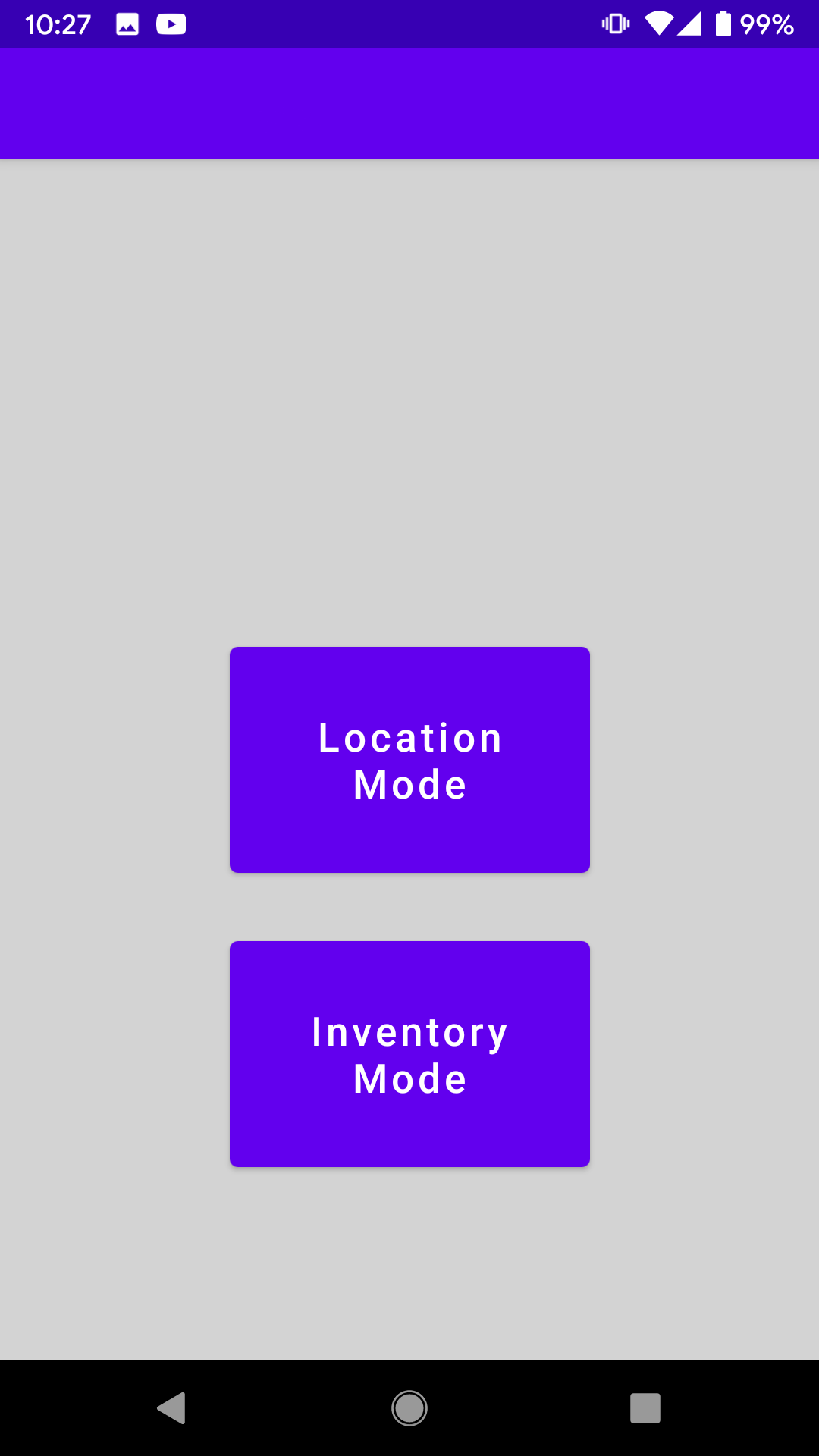}
\caption{\ApplicationName \hspace{0pt}'s home screen}
\label{fig:DIAL_Home_Page}
\end{subfigure}
\hfill
\begin{subfigure}[b]{0.26\textwidth}
\includegraphics[width=\textwidth]{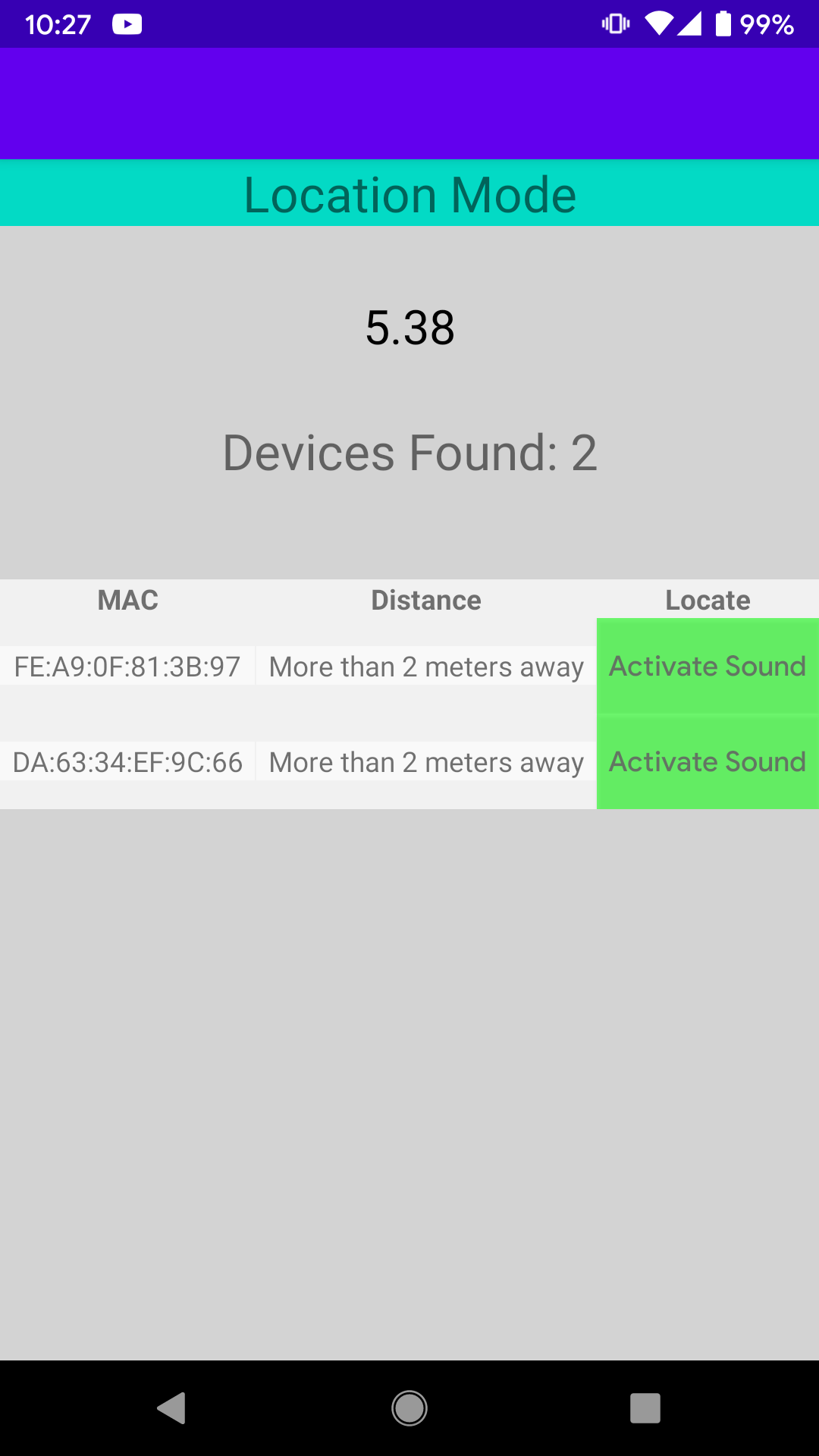}
\caption{\ApplicationName \hspace{0pt}'s location mode screen}
\label{fig:DIAL_Location_Mode}
\end{subfigure}
\hfill
\begin{subfigure}[b]{0.26\textwidth}
\includegraphics[width=\textwidth]{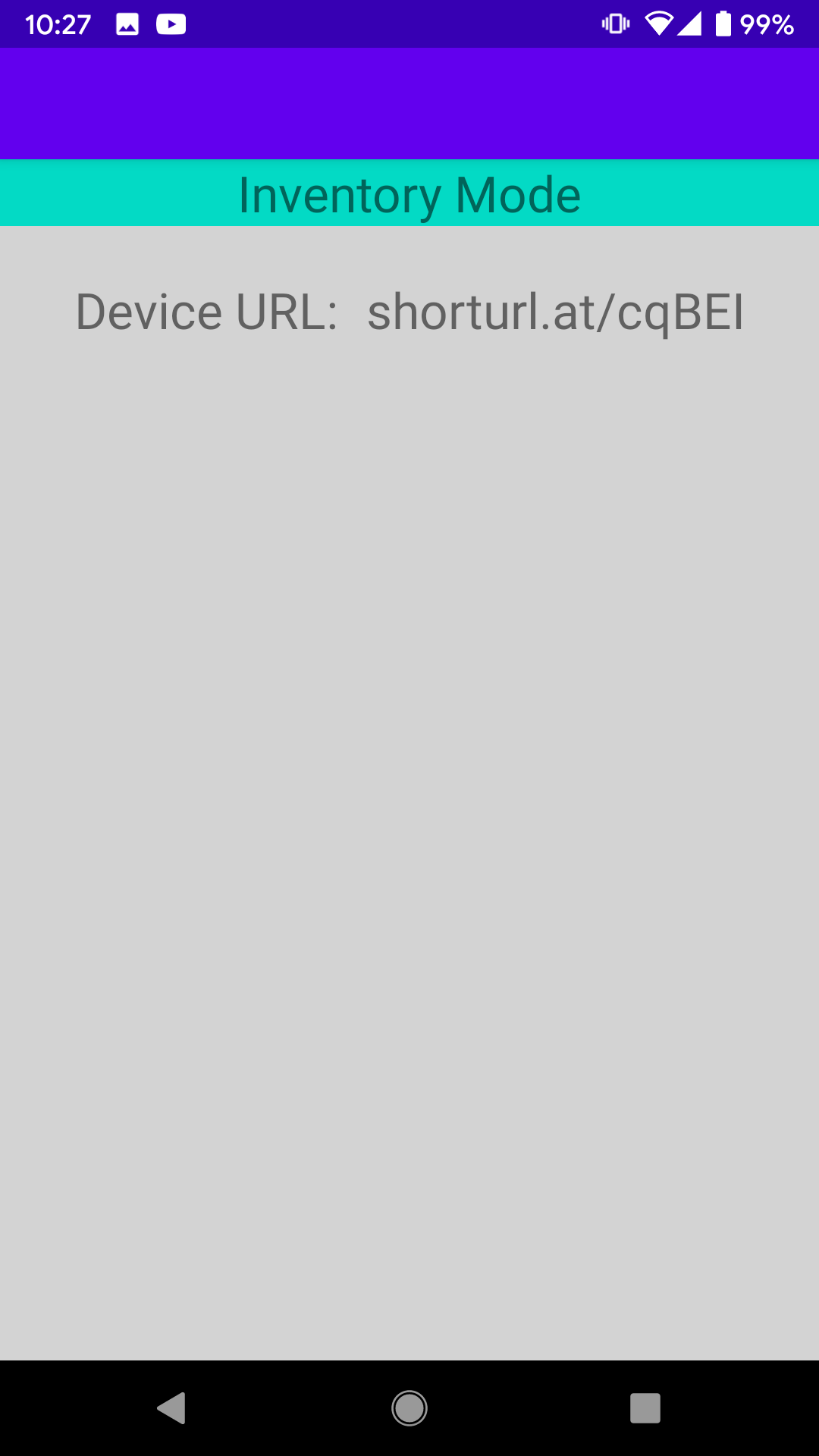}
\caption{\ApplicationName \hspace{0pt}'s inventory mode screen}
\label{fig:DIAL_Inventory_Mode}
\end{subfigure}
\caption{\ApplicationName's screens}
\end{figure*}

Besides the boards that serve as the microcontroller of each tag model, there are two other cheaper electronic components inside the tags. One such component is the \$0.95 buzzer that activates when the Adafruit boards receive a specifically formatted beacon \cite{mouser}. The other component is the coin-sized NFC card sold as 50 pieces from Walmart \cite{walmart.com}. Each card costs around \$0.76 and has a diameter of 25 millimeters \cite{walmart.com}. 

\subsection{\ApplicationName \hspace{0pt} Android Application}

The \ApplicationName \hspace{0pt} we developed greets users with its homepage that has two buttons. A home page snapshot is in Fig. \ref{fig:DIAL_Home_Page}.

The first button, labeled, ``Location Mode,'' redirects the user to another page containing all detected devices through their transmitted BLE beacons. Each discovered device will occupy one row of this list. Each row will have a button on its right side labeled ``Activate Buzzer'' or ``Activate Radar.'' The buzzer activation button will command the buzzer at the tag to play three different frequencies, three seconds long each. The radar activation button will take the user to an empty page with only the distance in meters printed on the screen. This distance will indicate how far the phone is from the tag. The location mode screen displaying tags is in Fig. \ref{fig:DIAL_Location_Mode}.

The second button, labeled, ``Inventory Mode,'' takes the user to the NFC page. This page activates the NFC reader. Once an NFC tag is in proximity, specifically within the centimeters range, the reader will print the information from the tag to the screen. This information may be a URL pointing to a web page for the attached device's data. An example of \ApplicationName \hspace{0pt} reading a shortened URL from the coin-sized NFC tag is in Fig. \ref{fig:DIAL_Inventory_Mode}.

\section{Evaluation}
We list all research questions as follows:
\begin{itemize}
  \item RQ1: What is the battery life of each tag?
  \item RQ2: What is the cost of each tag?
  \item RQ3: How quickly do users find the tags via \ApplicationName \hspace{0pt} in an apartment setting?
  \item RQ4: What is the user opinion on the usability of \ApplicationName?
\end{itemize}

This section will answer our research questions. We chose these metrics as they directly influence user convenience and the tags' adoption rate. We only evaluated the tags' discovery mode, as the inventory modes' underlying technology (i.e., NFC) did not have reliability issues in our experiments. Also, the information in the NFC is flexible depending on different design choices and privacy concerns. NFC also does not use a battery, and each NFC tag's cost is negligible.

We will then present our user study with \NumberOfParticipants \hspace{0pt} participants and deliver our SUS results. We did not stick the tags to sample IoT devices as it would not influence the results. Even if IoT devices use BLE and NFC, those protocols can reliably operate in environments with multiple transmitters. In addition, although our quarter-sized tags have three AA batteries attached to them, we expect them to fit with most IoT devices.

\subsection{RQ1: Battery Life of Each Tag}

\begin{table*}[ht]
\centering
\caption{Performance results from each tag model.}
\label{table:model_results}
\begin{tabular}{cccc}
\toprule
Model Name & Price & Battery Life Minimum (6000 mAh) & Battery Life Maximum (9000 mAh) \\ [0.5ex] %
\midrule
 \TagModelOne & \$21.66 & 250 days & 375 days \\
 \TagModelTwo & \$20.26 & 3.3 days & 5 days \\ [1ex]
\bottomrule
\end{tabular}
\end{table*}

We chose specific metrics to measure the quality of a tag. The first and most important metric for us is battery life. To provide users with the best experience, we wanted them to spend the least time with the physical tag. One possible scenario in which a user can interact with the physical tag is for maintenance reasons. The most common maintenance type we expect from users is battery replacement. 

The first design decision we made to reduce the maintenance time was to increase the battery capacity without drastically increasing its size. Most tags in the industry use coin batteries to keep the tags' size minimal. We assumed that increasing the battery life is more crucial than the tag aesthetics when attaching our tags to IoT devices. Thus, we chose three typical alkaline batteries in the standard AA size for our primary power source. They have a battery capacity of 2000 mAh to 3000 mAh with a cell voltage of 1.2 V to 1.5V \cite{jan_2010}. To calculate battery life, we divided the battery capacity by the current consumption of the tag. The result gave us the battery life in hours \cite{battery}. We divided the battery life metric into two sections: the minimum battery life, which assumes each battery has a capacity of 2000 mAh, and the maximum battery life, with a total of 3000 mAh \cite{jan_2010}. In our testing assumptions, we did not consider the board lifetime. We assume the boards function forever.

We also measured the current draw of the board via a standard USB power meter by plugging it into our computer. We connected the USB side of the micro-USB to the power meter and the other side to the board. Unfortunately, our power meter could only detect at least one milliampere. This limitation was problematic as we observed that the Adafruit ItsyBitsy nRF52840 Express board drew less than one milliampere of current when it both emitted and listened to BLE beacons. There are more expensive alternatives near \$5570 in the market that can measure currents at Femto levels \cite{testequity}. However, our power meter is much cheaper as it only costs approximately \$17 \cite{ebay}. 

We assume the Adafruit board has a current draw of 1 milliampere to circumvent this limitation and give an actual battery life. This assumption should provide us with an upper bound. Nevertheless, we found online posts discussing that the Adafruit board's current draw is less than one nano amperes when they transmit beacons with an interval of 100 milliseconds \cite{hathach_2019}. Measuring the current draw from the DWM1001-DEV board was more straightforward as the current draw was above one milliampere. The current draw for this board was 75mA when its UWB function was continuously running. Since the board in the \TagModelTwo \hspace{0pt} model was not power-optimized, our battery life for this solution is minimal \cite{nudel_2019}.

\subsection{RQ2: Tag Prices}

Another metric we used is the price of the solution. Price significantly influences consumers' purchasing behavior and sales \cite{han2001consumer}. Thus, we aimed to provide the cheapest solution. We mostly found this solution with devices aimed at hobbyists. Our research showed that vendors like Adafruit and Sparkfun give the most inexpensive boards. These vendors also use the Arduino IDE as their programmers. This practice is advantageous as Arduino IDE has an easy interface \cite{louis2016working, pan2018getting}. We also found the Arduino IDE to have the most documentation online compared to other board IDEs, such as SEGGER and Mbed. 

However, we have not found any UWB boards designed for hobbyists from these vendors. The cheapest option was the Qorvo DWM1001-DEV board that used SEGGER as its programming interface. Although SEGGER caused our process to slow down due to the scarcity of documentation, the DWM1001-DEV board still provided accurate results for us. After adding the buzzer and NFC card costs, each tag model price is in Table \ref{table:model_results}, combined with battery results from RQ1. 

\subsection{RQ3: Functionality of the Tags}

\begin{figure}[ht]
\centering
\includegraphics[width=0.5\columnwidth]{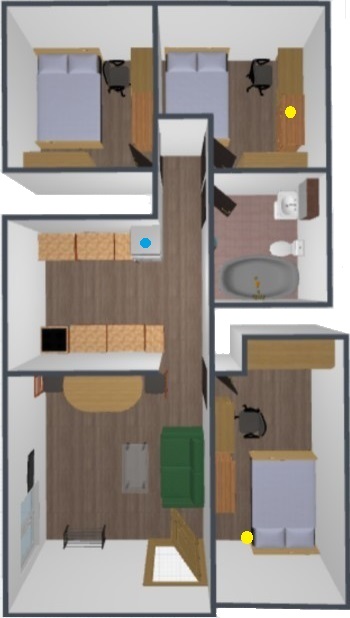}
\caption{The layout of the apartment and locations of the tags. The yellow dots indicate \TagModelOne \hspace{0pt} models, while the blue dot inside the refrigerator is the \TagModelTwo \hspace{0pt} tag. The yellow dot closer to the right bottom corner is \TagModelOne \hspace{0pt} 1, while the other one on the top-right corner is \TagModelOne \hspace{0pt} 2.}
\label{figure:home_layout}
\end{figure}

We hid two \TagModelOne \hspace{0pt} tags and one \TagModelTwo \hspace{0pt} tag inside an apartment. The apartment's layout and the tags' locations are in Fig. \ref{figure:home_layout}. While we put the \TagModelTwo \hspace{0pt} model inside the refrigerator to make it non-line-of-sight, we placed one \TagModelOne \hspace{0pt} tag, named \TagModelOne \hspace{0pt} 1, on top of the desk, while the other one, named \TagModelOne \hspace{0pt} 2, is on the floor next to the bed. Both tags were visible to users during the trials. In these trials, we named the \TagModelOne \hspace{0pt} tags 1 and 2 to distinguish each one.

We conducted these trials with \NumberOfParticipants \hspace{0pt} users whom \NumberOfParticipantsUndergraduates \hspace{0pt} were undergraduate students, and \NumberOfParticipantsGraduates \hspace{0pt} were graduates in various majors. \NumberOfParticipantsUndergraduatesSTEM \hspace{0pt} undergraduate students were in a STEM field, while \NumberOfParticipantsGraduatesSTEM \hspace{0pt} graduate students were in it too. We chose the users by asking random pedestrians near the University of Illinois Urbana Champaign campus to participate in our experiment and take our SUS survey in exchange for a doughnut. We explained that we hid three devices in an apartment they had not seen before and asked them to find them using \ ApplicationName's Location Mode. We gave them our Android phone with \ApplicationName \hspace{0pt} and timed them while they completed this task. 

Each participant started at the unit's entrance, at the bottom middle of Fig. \ref{figure:home_layout}. We first discussed the order in which they should find the tags since \ApplicationName \hspace{0pt} already had all tags on its screen as it was in their range. We asked them to locate the \TagModelTwo \hspace{0pt} tag, then the \TagModelOne \hspace{0pt} 1 and 2, respectively. Once the user found a tag, we reset the timer to record the next tag hunt. The average times of each trial and their standard deviations are in Table \ref{table:tag_hunt_trial_results}. It is also worth noting that we received IRB approval for this user study.

After the experiments, we examined our results in Table \ref{table:tag_hunt_trial_results}. Each trial has an average completion time of less than one minute, meaning that users spent less than a minute finding each tag. \TagModelOne \hspace{0pt} 1 and 2 had less than half the completion time of \TagModelTwo. However, \TagModelTwo \hspace{0pt} did not provide a direction toward itself. Thus, users had to experiment with moving around to find the direction where the distance decreased. This process prolonged the average trial completion time and negatively affected user convenience. However, since it is less than a minute, UWB-RAW is still a viable solution. Future iterations for this project can use the angle of arrival metric to give users a direction toward the tag.

\begin{table}[ht]
\centering
\caption{The tag hunt trial results: average time and standard deviation in seconds.}
\begin{tabular}{cccc}
\hline
& \TagModelTwo{} (s) & \TagModelOne \hspace{0pt} 1 (s) & \TagModelOne \hspace{0pt} 2 (s) \\
\hline
Avg. & 52.56 & 22.13 & 25.78 \\
Std. & 13.42 & 9.40 & 19.01 \\
\hline
\end{tabular}
\label{table:tag_hunt_trial_results}
\end{table}

\subsection{RQ4: Usability of the Tags}

\begin{table*}[h]
\centering
\caption{SUS survey score results.}
\begin{tabular}{lc}
\toprule
Question & Average Score \\ [0.5ex]
\midrule
I think that I want to use this system frequently. & 4.43 \\
I found the system unnecessarily complex. & 1.30 \\
I thought the system was easy to use. & 4.78 \\
I think that I would need the support of a technical person to be able to use this system. & 1.61 \\
I found that the various functions in this system were well integrated. & 4.74 \\
I thought there was too much inconsistency in this system. & 1.35 \\
I would imagine that most people would learn to use this system quickly. & 4.52 \\
I found the system very cumbersome to use. & 1.48 \\
I felt very confident using the system. & 4.74 \\
I needed to learn a lot of things before I could get going with this system. & 1.56 \\
\bottomrule
\end{tabular}
\label{table:sus_results}
\end{table*}

After our tag hunt trials, we asked the \NumberOfParticipants \hspace{0pt} users to rate their experience with \ApplicationName \hspace{0pt} and the tags as a whole system. We used the SUS template to design our survey. The SUS survey has ten questions about user experience; each question's answer is rated from one to five. 

Our SUS survey results in Table \ref{table:sus_results} were satisfactory. SUS questions' order determined the answer's scoring in SUS \cite{brooke1996sus}. Each question's answer is on a linear scale between one to five, with one strongly disagreed and five strongly agreed. Even-ordered questions' highest score was one, while their lowest score was five. Even-ordered questions are the second, fourth, sixth, eighth, and tenth questions, while the odd-ordered questions are the other ones. Even-ordered questions inquire about a negative aspect of the system. This inquiry method means a five would indicate a negative experience while a one would be positive. Odd-ordered questions had the highest score of five, while their lowest score was one. Odd-ordered questions would inquire about a positive aspect of the system. Thus, a five would indicate a positive experience.

We took the average of all the participants' answers to produce a final result for each question. Our overall SUS score is 89.77, while the highest SUS score is 100. SUS scores above 68 indicate above-average performance, while anything below 68 would be below average.

Our lowest-scored question is whether the user can interact
with this system without a technical person. Although there are lower-scored questions than 1.61 for this question, the highest score for even-ordered questions is one. So, the closer it is to five, the worse experience a user has. So, 1.61 would be equivalent to 4.39 for odd-numbered questions' reference. We suspect we have not included tutorials or self-explanatory tips in the mobile application. Future iterations can involve tip boxes and directions to solve this gap. Our second lowest-rated question is the first, with a 4.43, asking if the user would use this system frequently. We do not envision users interacting with the tags often, as they will perform device discovery only once when they move into a unit.

\section{Discussion}
This section will discuss the feasibility of the solutions and our future work for this project.

\subsection{Feasibility of our Tag Models: Price and Battery}

Here, we illustrate the feasibility of our tag models by comparing them with commercial options, Tiles, and AirTags because their technical specifications are easy to access. 

From the price perspective, the \TagModelOne{} model is a couple of USD cheaper than the \$25 Tile Mate \cite{tile}. Other Tile models, such as the \$35 Pro and the \$30 Sticker, are more expensive. In addition, alternative BLE breakout boards are cheaper than the Adafruit breakout board. The Nordic Semiconductor nRF52840-Dongle is only \$10 \cite{dongle}. We used the DWM1001-DEV board in the \TagModelTwo{} model because it was the cheapest and most user-friendly option that we could find. The Airtag price is \$29, while \TagModelTwo{} costs \$20.26 \cite{apple}.

From the battery consumption perspective, the \TagModelOne{} model has nearly a year of battery life. However, we found online posts stating that the battery consumption was under one nano ampere during beacon transmission with a 100-millisecond interval. We believe the life of the \TagModelOne{} model with replaceable AA batteries is much longer than one year. Tile models have a battery life of either three years with a non-replaceable battery or one year with a replaceable battery~\cite{tile}. The limitation for \TagModelOne{} is the lifetime of the ItsyBitsy microcontroller and its battery, as certain environmental conditions can cause electronics to degrade faster. 

The battery life of UWB ranges from 3.3 to 5 days as the DWM1001-DEV board is not optimized for power consumption \cite{nudel_2019}. Engineers designed it as an anchor point with an unlimited power supply \cite{nudel_2019}. Finding an alternative UWB board can be a challenging task. However, Airtags use UWB with their U1 chip \cite{apple}. Their solution can last longer than one year with a replaceable CR2032 coin cell battery \cite{apple}. Thus, a power-optimized UWB breakout board can last longer with three AA batteries. The U1 chip is not available for purchase, to our knowledge. Additional work is feasible for using U1 in our \TagModelTwo{} model to decrease power consumption. We expect that the \TagModelTwo{} model will be more practical given technical advancement in UWB that optimizes for battery in the future. 

In this paper, we did not conduct a user study for our tags' price acceptability, so we're unaware of our tags' attractiveness to landlords and manufacturers. An interview with landlords and smart home device manufacturers can improve our work by providing more concrete proof of price acceptability and an insight into how they can accept our tag models. We view this problem as one of the future works. 

\subsection{Future Work}

We only evaluated our tags in an apartment and shared space setting. Nevertheless, an evaluation in a generalized setting can be more informative to the community. A generalized environment would include buildings, such as large stadiums or skyscrapers. We can design a solution where a user could discover the tags without being in a room, like being on a farm. Several wireless protocols, such as Long Range (LoRa) or Zigbee, can prove helpful in a generalized environment. For instance, LoRa would be an excellent solution in grande buildings for tag discovery. A longer-range locating strategy with LoRa would be more convenient for users.

Two tag models to evaluate may not give readers a comprehensible understanding of solving device discovery and identification with wireless tags. There are other low-power wireless solutions designed for IoT systems. With the mentioned wireless protocols, we can build several other tag models useful in different scenarios, such as models for adverse environments or large buildings.

Our current models cannot accommodate some of these scenarios. For instance, we only use NFC for identification with centimeter ranges. If the IoT device the tag is attached to is in an adverse environment, such as inside the walls, it would be inconvenient for the user to identify it. Therefore, another protocol can be helpful. BLE can be a candidate technology, but the range is too extensive that neighbors could gain information regarding the hidden devices inside a unit. The Adafruit board and API allow programmers to configure the transmitter power to combat this. Observing the range of BLE in the lowest power configuration would be a worthwhile effort to monitor the feasibility of BLE in inventory mode.

The energy consumption of DWM1001-DEV is also not optimized, causing a battery life of only days \cite{nudel_2019}. We found that \TagModelTwo's UWB feature continuously runs in the example files we used \cite{decawave}. A more energy-saving method can be used, such as using the low-power BLE option to activate the UWB. The example files we use also have the nRF5 SDK. This powerful SDK tool allows developers to configure the UWB feature for various actions.

One last possibility for extending our work is to write a tutorial on what specific boards are needed and upload our example code to GitHub. The tutorial can guide users step-by-step on implementing the tag models if they need a customized solution.

\section{Related Work}
Researchers tackled the problem of device existence and location discovery in the literature using several tools. One such tool is PriView, which allows users to visualize nearby privacy-invasive devices via thermal cameras and VR \cite{prange2021priview}. Although some home appliances heat up when performing heavy loads, small devices like hidden sensors may only slightly heat. These sensors' heat may not be distinguishable if they are close to a larger appliance. Their VR solution only uses mockup solutions and does not have device detection implemented \cite{prange2021priview}. Some devices are out-of-sight, such as inside walls, and our solution covers these use cases. Fernandez et al.~\cite{fernandez2021privacy} utilized augmented reality to contextualize data disclosure and allow users to customize privacy filters on collected data. However, their AR interface is limited to two types of IoT devices and is restricted to line-of-sight, in which the user knows the device's existence in advance. Sharma et al.~\cite{sharma2022lumos} proposed Lumos, which enables users to identify and locate WiFi-connected hidden IoT devices and visualize their presence using an augmented reality interface. Lumos~\cite{sharma2022lumos} uses machine learning to tackle the challenge of identifying diverse devices with only limited features. 

Another solution comes from Song et al., as they used LED, WiFi, and mini speakers to have their participants locate their devices in their prototype \cite{song2020m}. Although this combination can help find and identify these IoT devices, WiFi is a power-intensive protocol, and they need to design their model with an unlimited power supply assumption. We focus on lowering the power consumption of our tags so users can attach them to any device that does not operate using the power grid. 

Several previous works focus on energy-efficient IoT device discovery. Chen et al.~\cite{chen2017energy} proposed a smart device discovery mechanism that adapts the scan window and the scan interval based on the number of redundant scanned devices within a scan window. Their mechanism reduces power consumption significantly compared to previous solutions, though it is limited to BLE devices. On the other hand, Sharma et al.~\cite{sharma2017energy} presented an energy-efficient architecture for device discovery in 5G-based IoT and Body Sensor Networks, i.e., BSNs, using uncrewed aerial vehicles, i.e., UAVs. However, the authors also pointed out several challenges surrounding the usage of UAVs \cite{sharma2017energy}.

\section{Conclusion}
We will explain the future trends of the device and identification problems and examine the future of some wireless protocols. Afterward, we will conclude this paper with final remarks.

\subsection{Future Trends}

As the number of IoT devices is increasing exponentially, more types of these devices will emerge in residential buildings \cite{oconnor2019homesnitch}. With this increase, new problems will occur, and the severity of existing issues will increase. One existing problem is device discovery and identification, as some IoT apparatuses can be malicious. A user should be aware of all the devices in his surroundings to protect his privacy. We expect researchers to leverage additional tools as solutions besides wireless. However, solving this problem via wireless protocols will provide a long-term solution.

It is also worth noting that researchers may develop additional wireless protocols that may be useful to the problem. WiFi HaLow is a tool that can be helpful in larger shared spaces, such as a stadium. It can operate on a cell battery for a year and has a one-kilometer range. 5G is also another wireless solution that can accommodate IoT devices. Our future work can use these tools to provide more capable tag models. 

\subsection{Final Remarks}

The primary purpose of the tags we developed is to allow more than a single user to interact with them to obtain the list of the devices in the shared space and learn about them. We created two tag models that use different wireless protocols: BLE and UWB. These protocols are for device discovery and identification. We use NFC in both tag models for device information extraction by users only in proximity. An Android application, \ApplicationName \hspace{0pt}, also allows the user to interact with the tags, specifically seeing every tag in the room as a list and options to find tags via its buzzer or UWB. This solution has not been done before for device discovery. 

Relevant solutions we found for this problem are tools, such as thermal cameras, LEDs, or VR \cite{song2020m,prange2021priview}. However, each solution has issues, such as LEDs needing to be in sight, and our tagging solution is robust to these errors. Therefore, we believe using BLE and UWB to solve this problem is worthwhile.

\begin{acks}
We would like to thank Hyun Bin Lee for participating in our discussion and finding related papers to our project. His insights were very valuable and helped us understand the field.
\end{acks}

\bibliographystyle{ACM-Reference-Format}
\bibliography{sample-base}

\end{document}